\documentclass[12pt]{article}
\usepackage{graphics}
\usepackage[]{epsfig}	       
\begin{document}
 
\thispagestyle{empty}
 
\title{Interacting electrons in a magnetic field in a center-of-mass free basis.}
\author{
Peter Kramer, Institut f\"{u}r Theoretische Physik,\\ 
Universit\"{a}t T\"{u}bingen, Germany\\
Tobias Kramer, Mads-Clausen Institute\\University of Southern Denmark, S{\o}nderborg, Denmark}
\maketitle

\begin{abstract}
We present an extension of the spin-adapted configuration-interaction method for the computation of four electrons
in a quasi two-dimensional quantum dot.
By  a group-theoretical decomposition of the basis set and working with  relative and center-of-mass coordinates we 
obtain an analytical identification of all spurious center-of-mass states of the Coulomb-interacting electrons.  
We find a substantial reduction in the basis set used for numerical computations.
At the same time we increase  the accuracy compared to the standard spin-adapted configuration-interaction method (SACI)
due to  the absence of distortions caused by an unbalanced cut-off of center-of-mass excitations.
\end{abstract}

Dedicated to Margarita and Vladimir Man'ko on the occasion of their 150th birthday.

\section{Introduction.}

The  Schr\"odinger  equation  of an  interacting many-body system can be solved analytically only for specific interaction potentials and for a very restricted number of particles.
A wide range of approximation methods has been developed to determine the ground and excited states of many-body systems found in nuclear, atomic, and condensed matter systems.
Here, we focus on quasi two-dimensional electronic systems, which are experimentally realized in quantum-dots.
Quantum dots are often embedded in layered semiconductor structures. The strong confinement along the vertical direction perpendicular to the layer suggests an effectively two-dimensional description along the remaining two lateral dimensions.
Besides the electron-electron interactions, an additional external confinement potential along the lateral directions is created by etching and gating of the semiconductor device.
Quantum dots can be regarded as {\em artificial atoms}, albeit with a  central potential different from the nuclear Coulomb attraction. Their excitation spectra are probed by electronic and optical measurements, also under the influence of additional magnetic fields, which gives rise to a Landau-level structure. 

With the discovery of the fractional Quantum Hall effect (FQHE) in systems of interacting electrons in quasi two-dimensional systems, 
many theoretical approaches for studying interacting electrons have been proposed, including the celebrated Laughlin wave-function for electrons in the lowest Landau level.
In the experimentally realized Hall devices translational symmetry is broken by the current source and drain contacts, which lead to the formation of hot-spots with high electric field values \cite{Kramer2009c,Kramer2009b}.

A comparison of the various theoretical approaches with numerical methods is often performed for few-electron quantum dots.
Analytic \cite{Vercin1991,Taut1994} and numerical \cite{Wagner1992} solutions for the two electron case show the interplay of Coulomb interaction and confinement potential, leading to alternating spin polarization of the ground state in a quantum-dot as function of magnetic field \cite{Su1994}.
By separation of the two-electron case into the center-of-mass motion and the relative part, semiclassical solutions have been constructed \cite{Klama1998a} and compared to the exact solutions \cite{Kramer2010c,Grossmann2011,Escobar2014}.
For three electrons, the relative-coordinate basis set and expressions for the Coulomb matrix elements are given in \cite{Simonovic2006}.

In the popular {\em configuration interaction} (CI) approach a diagonalization of the Hamiltonian including the Coulomb-interaction is done with respect to a product basis of single-particle Slater determinants.
Convergence is generally checked by systematically enlarging the size of the basis set, for instance by increasing the value
of the cut-off energy up to which all states are included for the non-interacting system \cite{Tavernier2003a}.

With increasing number of basis states, the computation of all interacting matrix-elements and the final diagonalization of the Hamiltonian quickly exhausts computer resources. Thus a reduction of the basis set while maintaining accuracy is required for studying the excitation spectra of few body systems.

If the reduction is based on underlying group-theoretical properties, an exact decomposition of the basis set into irreducible blocks is possible. A diagonalization of the smaller blocks is more easily achievable and the decomposition provides physical insight into the system.

One example is the separation of the spatial part of the wavefunction from the spinor states for electrons. 
If we require that the product of spatial and spin-part represents electrons (antisymmetric states), the basis
set is decomposed into separate spin states.
This results in spin-adapted (SA) configuration methods, for instance implemented in the SACI program code \cite{SACI,SACI2}.

At this point  SACI methods used in condensed matter theory for more than 3 electrons stop and do not exploit additional properties of the basis set, such as a decomposition into relative and (interaction-free) center-of-mass ({\em cm}) states.
The separation of the basis set into a product state of relative and center-of-mass part is of particular use if the Hamiltonian of the $n$-electron system is given as a sum of operators acting on center-of-mass and relative coordinates respectively.
For electrons confined to a two-dimensional plane perpendicular to a uniform magnetic field this separation of the Hamiltonian is possible as long as the confinement potential in the two-dimensional plane does not lead to an admixing of relative and center-of-mass coordinates \cite{Kohn1961,Avron1978}. 
For molecular or atomic systems moving in three dimensions across uniform magnetic fields, additional couplings between relative and center-of-mass coordinates occur, i.e.\ the motional Stark effect \cite{Johnson1983,Yukich2003,Escobar2014}.
The importance of a systematic separation of relative coordinates from the center-of-mass excitation for studying correlated two-dimensional electron systems for three electrons was pointed out early by Laughlin \cite{Laughlin1983a} and noticed recently again for more than three electrons  \cite{Yannouleas2010a}. 
Laughlin's trial wave function  by construction is expressed only in relative coordinates and can be interpreted and numerically handled similar to a classical plasma of interacting particles \cite{Kramer2012a}.

We note that both in many-electron and nuclear physics, the interactions between the particles  are exclusively functions of relative coordinates. Then the same shift vector applied to any single particle coordinate does not change the hamiltonian.
By  the well-known Noether theory of symmetry it then follows that the total momentum ${\bf P}$ of the system is conserved.
For an unusual but powerful example of a conserved quantity we refer to \cite{KR09}.
Relative  coordinates, in contrast to the single particle coordinates, are free of the $cm$ coordinate. In quantum mechanics,
the components of ${\bf P}$ are proportional to the partial derivatives of the $cm$ coordinate. Therefore,
if the  states depend on   relative coordinates only, we have ${\bf P=0}$. Another possible option which we shall adopt  is to 
demand the $cm$ state to have zero oscillator excitation. The paradigms of  nuclear theory show \cite{KM66,KM68b,AKM69} that  relative coordinates are the natural tools  to understand and compute nuclear  
scattering and reactions. In nuclear structure, such coordinates may be adapted and serve to describe few-particle systems \cite{KM68a,AKM69}, clustering, and decay processes. The construction of orbital states can be  achieved by use of 
group and representation theory \cite{KM68b,K74}. The orbital states are coupled with spin states to antisymmetric states. The adaption to orbital permutational symmetry for composite systems in relative coordinates becomes more involved, but was extensively studied in \cite{KJS80}.

We report here work on 2D electrons in an external perpendicular magnetic field, 
so that the 2D total momentum is still conserved \cite{Kohn1961,Avron1978}. 
We use harmonic oscillator states and the methods of Bargmann Hilbert space. 
The electron-electron Coulomb repulsion is computed in matrix form and included into the diagonalization.
For $n=4$ electrons we  use the relative tetrahedral coordinates from \cite{KM66} with simple properties under permutations. 
The spin states for total spin $S=2,1,0$ are explicitly determined by projection operators.\\
Single-particle coordinates are used to construct antisymmetric states as Slater determinants of single-particle states.
If these states are taken as harmonic oscillator states, Slater determinants cannot avoid the intrusion of $cm$ excitations 
into state space which give raise, in addition to the proper states,  to so-called spurious states of spurious angular momentum. 
The true role of spurious states  can  be seen only by the parallel use of relative and $cm$ coordinates.
We shall present in section 7 in particular for total spin $S=0$, a complete analysis of these spurious states. 
Their intrusion of state space is demonstrated in the spectrum of determinantal eigenstates of the standard SACI approach \cite{SACI,SACI2,Tavernier2003a}. 
Moreover, if determinantal  eigenstates are 
variationally computed in a subspace defined by a fixed maximal oscillator excitation, we find that the admixture of spurious states 
distorts this subspace and the variational results. In our proper treatment, any eigenstate in relative coordinates 
is the basis  to a set of equidistant spurious states of increasing $cm$ excitation. 

\section{Total spin S and orbital partition f.} 
The total spin of $n$ electrons ranges from the maximum $S=n/2$ to the minimum value,
$S=1/2$ for $n$ odd and $S=0$ for $n$  even. The permutational symmetry of the spin state is characterized by a partition $f^S=[f_1^Sf_2^S],\: f_1^S+f_2^S=n,\: f_1^S \geq f_2^S$ of two rows 
$[f_1^S,f_2^S]$ only, because there are for any electron  only the two spin (up, down) states .  The total spin $S$ is related one-to-one  to the spin partition $f^S$ by 
\begin{equation}
\label{es1}
S(f^S)=\frac{1}{2}(f_1^S-f_2^S). 
\end{equation}
Since the electrons are fermions, the combined orbital plus spin  state must be antisymmetric with partition $f= [11..1]$. Then is follows that the only allowed orbital partition 
$f$ is associate to the spin partition $f^S$, that is,  its Young diagram consists of $n$ square boxes, arranged in $f_2^S$ rows  
of length 2, and $(f_1^S-f_2^S)$ rows of length 1. To the orbital partition $f$ there belong orthogonal partner functions whose number we denote by $|f|$. They may be labelled by Young tableaus $r,s...$. Any Young tableau is a  filling of the boxes of the Young diagram $f$ by the numbers $1,2,...,n$, such   that the numbers increase both to the right and downwards. The application of 
permutations to an orbital  state of $n$ particles, denoted as $|\alpha^n fr\rangle$, is completely determined by the tableau $r$.
The antisymmetric fermion state, with respect to permutations, takes the form of a sum of products of orbital and spin states,
\begin{equation}
 \label{es2}
 |[1^n]\alpha^n  S \rangle= |f|^{-\frac{1}{2}} \sum_r |\alpha^n f r\rangle |S f^S r^S\rangle,
\end{equation}
where $r^S$ is the Young tableau of the spin partition $f^S$, determined  from the orbital tableau $r$ by reflection in the main diagonal of its orbital Young diagram.  

The operators $O$ for the external magnetic field, the central field  and the interactions are all invariant under permutations and independent of the spin. 
Since the spin basis states $|S f^Sr^S\rangle$ are orthogonal, the matrix elements of all spin-independent  
operators $O$ between antisymmetric states reduce to the  spin-free sum of orbital matrix elements 
\begin{equation}
 \label{es3}
\langle 1^n S|O|1^n S\rangle 
= |f|^{-1} \sum_r \langle \alpha^n f r|O|\alpha^n f r \rangle. 
\end{equation}
Here the total spin $S$ is fixed from eq.~\ref{es1} by half the difference of the column length of the 
orbital partition $f$. From eq.~\ref{es3} it follows that we never need the spin states
but must compute $|f|$ orbital diagonal matrix elements. 

{\em Example}: For $n=4$ particles, the possible values of the total spin are 
$S=2,1,0$. The corresponding spin partitions are $f^S=[4],[31],[22]$, and the orbital partitions are $f= [1^4],  [211],[22] $  of dimensions $|f|$ =1, 3, 2 respectively.

So we need the construction of $|f|$ orbital states with 
an orbital partition $f$, fixed by the total spin $S$,  and by its  Young tableaus $r$. 

\subsection{Orbital Young diagrams and representation of transpositions.}
To project states of orbital symmetry $f$ with the Young tableau $r,s$, we use 
the matrix elements of the representation $D^f_{r,s}(p)$ which are determined by
the tableaus.  A Young diagram is an arrangement  of square boxes in rows of length $f_1f_2...f_n$. In a Young tableau $r$, we label  these  boxes  by $1,2,...,n$ such that the numbers increase to the right and downwards.
There are exactly $|f|$ Young diagrams for the partition $f$.
The row and column of the box $i$ are then  $(\alpha_i,\beta_i)$. For a pair of boxes 
$(i,j)$ we define the axial distance \cite{KJS80} p.28 by
\begin{equation}
 \label{es10}
 \tau_{i j}= (\alpha_i-\alpha_j)-(\beta_i-\beta_j).
\end{equation}
We first determine the matrices of the representation for the $(n-1)$ transpositions 
$p= (1,2),(2,3),...,(n-1,n)$ of $S(n)$. These transposition by multiplication generate 
all elements of $S(n)$, and so all representation matrices can be built by matrix multiplication from the representations of these transpositions.
For the permutation which interchanges $i$ with $j=i+1$ and the Young tableau $r$,
there may be a second tableau $s$ with boxes $(i,j)$ interchanged.
If there is no such second tableau, the two boxes must appear in succession in the same row or the same column of the Young tableau. In these two cases 
the matrix element of the representation are given by 
\begin{equation}
\label{es11}
D^f_{r s}(i,i+1)= 1/\tau_{i,i+1}= \pm 1. 
\end{equation}
Otherwise, the non-trivial  part of the representation is given by the $2 \times 2$ matrix 
\begin{equation}
 \label{es12}
d^f_{r,s}(i,i+1)=
\left[
\begin{array}{ll}
1/\tau_{i,i+1}& \sqrt{1-(\tau_{i,i+1})^{-2}}\\
\sqrt{1-(\tau_{i,i+1})^{-2}}& -1/\tau_{i,i+s}\\
\end{array}
\right].
\end{equation}

The operator of the transposition $T(p)=T(a,b)$ interchanges  the coordinates of the single particle indexed by the number $a$ 
with those  of the particle indexed by $b$.
Following the Wigner \cite{WI59} pp.102-6 prescription, to assure a homomorphic action we must associate to any permutation $p$ the inverse
action $T(p)$ of the permutation on the particle indices.

The Young operators \cite{KJS80} pp. 25-26 are given by 
\begin{equation}
 \label{es13}
 c^f_{rs}=\frac{|f|}{n!} \sum_{p\in S(n)} D^f_{r,s}(p)  T(p),\: [c^f_{rs}]^{\dagger}= c^f_{sr}.
\end{equation}
These operators must be applied to the highest polynomials eq.~\ref{es21} of fixed pseudo angular momentum $\Lambda$ (see Sect.~\ref{sec:pseudo}), provided that from \cite{KM66} we know that $\Lambda$ can yield the orbital partition $f$. 
If a Young operator gives zero when applied to the polynomial, we first apply a lowering operator eq.~\ref{es23}
and then again try the Young operator. Exhausting all values of the pseudo angular component we must find 
all the states of orbital symmetry.

For an antisymmetric state eq.~\ref{es2} we use orbital transposition operators $T(s,t)$ to write
\begin{eqnarray}
\label{es13e}
&&\sum_{s<t} V(s,t) = \sum_{s<t} T(s,1) T(t,2) V(1,2)T(t,2)T(s,1),
\\
&&\langle \alpha^n [1^n]S|\sum_{s<t} V(s,t)|\alpha^n[1^n]S\rangle 
\\ \nonumber
&& = \frac{1}{|f|}\sum_{s<t}
\sum_r \langle f r|T(s,1) T(t,2) V(1,2)T(t,2)T(s,1)|f r\rangle
\\ \nonumber 
&&=\frac{1}{|f|} \sum_{s<t} \sum_{r}\sum_{r',r''}  \langle fr| T(s,1) T(t,2)|fr'\rangle 
\langle f r'|V(1,2)|f r''\rangle \langle fr''|T(t,2)T(s,1)|fr\rangle
\\ \nonumber 
&& = \frac{n(n-1)}{2} \frac{1}{|f|}\sum_r \langle f r|V(1,2)|f r\rangle. 
\end{eqnarray}
The third  line of eq.~\ref{es13e} follows from the orthogonality of the spin states in eq.~\ref{es2}.
In   the last line we use the involutive property of transpositions and the orthogonality of Young tableaus,

\begin{eqnarray}
\label{es13x} 
&& [T(s,1)]^2=[T(t,2)]^2=1,  
\\ \nonumber
&&\langle f r''|fr'\rangle 
= \delta_{r'',r'}.
\end{eqnarray}
{\bf Prop}: The matrix element of orbital twobody interactions, invariant under permutations, between antisymmetric states of fixed total spin $S$, can be reduced to a sum over diagonal orbital matrix elements of the interaction of the  first pair between  states characterized by Young tableaus , multiplied by the number of pairs.

To construct states with other tableaus of the same partition, we provide a ladder procedure  by employing 
the properties of the representation. Given an initial basis function 
$|fr\rangle$, we know from eq.~\ref{es12} that the permutation $(i,i+1)$ connects it at most to a 
second basis function  $|fs\rangle$ whose Young tableau  differs  by 
an interchange of the numbers $(i,i+1)$ . We rewrite eq.~\ref{es12} in the form 
\begin{eqnarray}
 \label{es14}
&&  T(i,i+1)|fr\rangle = |fr\rangle \tau^{-1} + |fs\rangle \sqrt{1-(\tau)^{-2}}, 
 \nonumber  \\
&& |fs\rangle =|fr\rangle \frac{-\tau^{-1}}{\sqrt{1-(\tau)^{-2}}} 
+ T(i,i+1)|fr\rangle \frac{1}{\sqrt {1-(\tau)^{-2}}}. 
\end{eqnarray}
Therefore once we know how to apply $T(i,i+1)$ to the state $|fr\rangle$, we can 
construct $|fs\rangle$ as the linear combination eq.~\ref{es14}, second line.
With the operator eq.~\ref{es13} we have constructed the state with the highest Young  tableau. 

{\bf Prop}: With steps as in eq.~\ref{es14} we can generate all other Young tableaus from an initial one. 

Examples are given in the following subsections.

\subsection{Construction of permutations.}

In an orbital Young operator eq.~\ref{es13}, the sum runs over all permutations of $S(n)$. 
This sum may be broken into classes $k$, characterized by $i_1, i_2,..,i_n$ independent and commuting cyclic permutations  of length $1, 2, ...,n$ where 
\begin{equation}
 \label{es14a}
1i_1+2i_2+...+n i_n=n. 
\end{equation}
Any cyclic permutations in turn can be factorized into transpositions as
\begin{equation}
\label{es14b}
(abcd...)= (ab)(bc)(cd)(..)...
\end{equation}
and any transposition is conjugate to one of the $n$ generators $(12),(23),...$ of $S(n)$, see eq.~\ref{es28}.
By the representation $p \rightarrow D(p)$, all the factorizations of permutations are converted into matrix factorizations.
A systematic enumeration of all $n!$ permutations can go in terms of classes $k$ of conjugate elements.
The number of permutations in the class $k$ corresponding to eq.~\ref{es14a} according to Weyl \cite{WE30} p.329 is
\begin{equation}
 \label{es14c} 
n(k)= \frac{n!}{1^{i_1}i_1!2^{i_2}i_2!...}.
\end{equation}

\section{Orbital states, unitary quantum numbers, total angular momentum.}
For a single electron in 2D, we label its state by the oscillator excitation $\nu$, by 
the numbers $n^+,  n^-, n^+ +n^-=\nu$ of right and left circular oscillator quanta, and by the angular momentum $l=n^+-n^-$. 

Note the following relations for coordinates and derivatives of a single particle in 2D, passed to  a circular setting:
\begin{eqnarray}
\label{es14d}
&& z^+:=\sqrt{\frac{1}{2}}(z^1+iz^2),\:  z^-:=\sqrt{\frac{1}{2}}(z^1-iz^2),
\\ \nonumber
&& z^1=\sqrt{\frac{1}{2}}(z^++z^-),\:  z^2=-i\sqrt{\frac{1}{2}}(-z^++z^-),
\\ \nonumber 
&& \partial^+=  \sqrt{\frac{1}{2}}(\partial^1-i\partial^2),\:
\partial^-=\sqrt{\frac{1}{2}}(\partial^1+i\partial^2),
\\ \nonumber 
&& \partial^1=\sqrt{\frac{1}{2}}(\partial^++\partial^-),\:
\partial^2=  i\sqrt{\frac{1}{2}}(\partial^+-\partial^-),
\\ \nonumber
\end{eqnarray}
and the quadratic relations
\begin{eqnarray}
\label{es4e}
&& z^1\partial^1+z^2\partial^2= z^+\partial^++z^-\partial^-
\\ \nonumber
&& z^1z^1+z^2z^2=z^+z^-+z^-z^+.
\end{eqnarray}
The last relations are easily extended to scalar products of  several vectors, labelled and contracted wrt lower indices. These yield generators of 
the symplectic group $Sp(2,R)$.

The states of $n$ electrons are then labelled by the summed up number $N$ of oscillator quanta and by the summed up angular momentum 
\begin{equation}
 \label{es15}
L= \sum_{s=1}^n (n^+_s-n^-_s).
\end{equation}
It is desirable to find other (integer) orbital quantum numbers which if possible can label at least in part 
an orthogonal set of orbital states. For this purpose 
we use the scheme of \cite{KM66} whose quantum numbers arise 
from group representations. The unitary group of all orbital 
degrees of freedom is the  unitary group $U(2n)$, its commuting subgroups acting on the 2 orbital or the 
$(n-1)$ relative motion degrees of freedom are $U(2)\times U(3)$. 
Their representations are both given by the same two-component integer partitions $[h_1h_2], h_1+h_2=N,\: h_1\geq h_2\geq 0$.
The representation of the subgroup $SO(2)<U(2)$ is characterized by the angular momentum $L$.
For given partition $[h_1h_2]$ the angular momentum ranges from its maximum 
$L=h_1-h_2, h_1-h_2-2... $ to its  minimum value $1$ or $0$ for $(h_1-h_2)$ odd or even  respectively.

\subsection{Lowest Landau level $LLL$.}
In the 2D magnetic field, the lowest Landau level $LLL$ arises if the summed up angular momentum 
$L$ equals the summed up excitation $N$. In terms of the partition $[h_1h_2]$ for the unitary group $U(2)$,
we have $N=h_1+h_2$. The weights, i.e. the degrees $(w_1,w_2)$ of the vectors indexed $1,2$, are restricted to the 
range $h_1\leq w_i\leq h_2$, and the maximum angular momentum is $L={\rm max} (w_1-w_2)=h_1-h_2$. 
The lowest Landau level therefore enforces $h_1-h_2=h_1+h_2, h_2=0$.
This condition selects  among the partitions of $N$ the single case 
$[h_1h_2]=[N0], L=N$. Only for this partition can the excitation $N$ be carried by a single vector.
For higher Landau levels this condition is relaxed and new unitary partitions are allowed.

\subsection{Pseudo angular momentum and lowering operators.}\label{sec:pseudo}

The group $U(3)$ has an orthogonal subgroup $SO(3)$ whose representation label  we term the 
pseudo angular momentum $\Lambda$. The possible values of $\Lambda$ were given in 
\cite{BM62}. 
Here the authors use the subgroup chain 
\begin{equation}
 \label{es16}
U(3)>SU(3)>SO(3), 
\end{equation}
with branching of representation labels 
\begin{equation}
 \label{es17}
[h_1h_2]> (\lambda',\mu')> \Lambda,\: (\lambda',\mu')=(h_1-h_2, h_2). 
\end{equation}
The range of these labels is determined by specific branchings from the representation in the 
next higher subgroup which we shall discuss in more detail.

We treat the oscillator states by the methods of Bargmann Hilbert space of analytic functions \cite{KJS80}.
We specifically note that in what follows we must {\em interchange the role of the groups and subgroups} $U(2), U(3)$
compared to \cite{BM62}, and {\em convert angular into pseudo angular momentum.}
This allows us to take advantage of the  group theory given in \cite{BM62}, but in a new and different interpretation.

By {\em upper indices} $j=1,2$ (or upper $\pm$), see below, we denote the Cartesian vector components in 2D, and by 
{\em lower Cartesian indices} $s=(1,2,3)$ or spherical indices  $(+,-,0)$ 
three (or more) relative vectors of the four (or more) particles.
The normalized states of angular momentum $LM$ of a single particle with oscillator excitation $N$ in the Bargmann representation are
\begin{equation}
 \label{es18}
P^N_{LM}(z)= A_{NL}(z\cdot z)^{\frac{1}{2}(N-L)}Y_{LM}(z),\: 
A_{NL}=(-1)^{\frac{1}{2}(N-L)}[\frac{4\pi}{(N+L+1)!!(N-L)!!}]^{\frac{1}{2}}.
\end{equation}
with $Y_{LM}(z)$ a solid spherical harmonic \cite{ED57} p.69.
The coordinates may then be taken as the six linear combinations of Cartesian coordinates
\begin{eqnarray}
\label{es19}
&&z^+_+= \sqrt{\frac{1}{2}}[(z_1^1+iz_2^1)+ i(z_1^2+iz_2^2)],
 \\ \nonumber 
&&z^+_-= \sqrt{\frac{1}{2}}[(z_1^1-iz_2^1)+ i(z_1^2-iz_2^2)],\: z^+_0=z^+_3,
 \\ \nonumber 
&&z^-_+= \sqrt{\frac{1}{2}}[(z_1^1+iz_2^1)- i(z_1^2+iz_2^2)],
 \\ \nonumber 
&&z^-_-= \sqrt{\frac{1}{2}}[(z_1^1-iz_2^1)- i(z_1^2-iz_2^2)], \: z^-_0=z^-_3.
 \\ \nonumber 
\end{eqnarray}
The lower Cartesian indices $(1,2,3)$ denote three relative coordinates like the Jacobi or 
tetrahedral ones. The permutations of particles act on these coordinates and introduce linear 
combinations wrt the lower indices.

Next we introduce  from \cite{BM62} p.187 for $n=4$ the elementary polynomials
\begin{eqnarray}
\label{es20}
&\eta_+:= z_1^1+iz_2^1,
\\ \nonumber
&v_+:=i(z_0^1z_+^2 -z_0^2z_+^1),
\\ \nonumber 
&w_+:=[(z^1\times z^2)\times z^1]_+
\\ \nonumber
&s:=z_+^1z_-^1+ z_0^1z_0^1,
\\ \nonumber 
&t:= (z^1\cdot z^1)(z^2\cdot z^2)-(z^1\cdot z^2)^2,
\\ \nonumber 
\end{eqnarray}
In \cite{BM62} p. 187 we find in the scheme eq.~\ref{es16} the linearly independent polynomials 
\begin{eqnarray}
\label{es21}
&h_1-\Lambda \:{\rm even}:
\\ \nonumber
&P_{h_1h_2\Lambda}=\eta_+^{(\Lambda-h_2+2q)}\: v_+^{(h_2-2q)}\:s^{((h_1-\Lambda)/2-q)}\:t^q,
\\ \nonumber 
&0\leq 2q \leq h_2,
\\ \nonumber 
&h_2-\Lambda\leq 2q\leq h_1-\Lambda
\\ \nonumber
&h_1-\Lambda \: {\rm odd}:
\\ \nonumber
&P_{h_1h_2\Lambda}= w_+\: \eta_+^{(\Lambda-h_2+2q)}\:v_+^{(h_2-1-2q)}\:s^{((h_1-\Lambda-1)/2-q)}\:t^q,
\\ \nonumber 
&0\leq 2q \leq h_2-1,
\\ \nonumber 
&h_2-\Lambda\leq 2q\leq h_1-1-\Lambda.
\end{eqnarray}
Here all numbers in exponentials must be non-negative, and $q$ is an integer which labels the multiplicity of a repeated pseudo angular momentum $\Lambda$ 
for a given partition $[h_1h_2]$ of $U(3)$. These multiple states of equal pseudo angular momentum are  indexed by $q$ and linearly independent, but not yet orthogonal.
Orthogonality must be  achieved by standard matrix methods.
The permutation group $S(4)$ acting on relative coordinates is a subgroup of the pseudo rotation group  
$SO(3)$. 
The multiplicity of its representations, see Table~\ref{table1}, was given for all orbital partitions $f$ of $4$ particles in \cite{KM66}. The polynomials eq.~(\ref{es21}) have the maximum component $\Lambda$ of the pseudo angular momentum.
Lowering of the component is achieved by applying  
the first order lowering differential operator
\begin{eqnarray}
\label{es22}
& \Lambda_-=\Lambda_1-i\Lambda_2= z_-^1{\partial_0^1}-2z_0^1{\partial}_+^1+z_-^2{\partial}_0^2-2z_0^2{\partial}_+^2,
\\ \nonumber 
&\Lambda_3=z_+^1{\partial}_+^1-z_-^1{\partial}_-^1 +z_+^2{\partial}_+^2-z_-^2{\partial}_-^2. 
\end{eqnarray}

Finally we can also lower the angular momentum $L$ from its maximum value $L=(h_1-h_2)$ in steps of $2$ by application of the
angular momentum operators  
\begin{eqnarray}
\label{es23}
&& L^3= z_1^+{\partial}_1^+ -z_1^-{\partial}_1^- +z_2^+{\partial}_2^+-z_2^-{\partial}_2^-+z_3^+{\partial}_3^+-z_3^-{\partial}_3^-,
\\ \nonumber
&&L^-= z_1^- {\partial}_1^+ +z_2^-{\partial}_2^+ +z_3^-{\partial}_3^+,
\\ \nonumber 
&&L^+= z_1^+{\partial}_1^- +z_2^+{\partial}_2^- +z_3^+{\partial}_3^-
\end{eqnarray}

\subsection{From pseudo angular momentum $\Lambda$\\ to orbital partitions $f$.}

\begin{table}
$
 \begin{array}{|l|lllll|l|l|} \hline
 \Lambda^-& [1^4] &[211]& [22]  &[31] &[4]  & \Lambda^{\pi}[1^4]&\Lambda^{\pi}[4]\\ \hline
 \Lambda^+& [4]   &[31] &[22]   &[211]&[1^4]&  &\\ \hline 
 0        & 1   &   &   &   &   & 0^-&0^+\\
 1        &     &   &   &1  &   &   &\\
 2        &     &1  &1  &   &   &   &\\
 3        &     &1  &   &1  &1  &3^+&3^-\\
 4        & 1   &1  &1  &1  &   &4^-&4^+\\
 5        &     &1  &1  &2  &   &   &   \\
 6        & 1   &2  &1  &1  &1  &6^+6^-&6^-6^+\\
 7        &     &2  &1  &2  &1  &7^+&7^-\\
 8        & 1   &2  &2  &2  &   &8^-&8^+  \\
 9        &1    &2  &1  &3  &1  &9^+9^-&9^-9^+\\
 10       &1    &3  &2  &2  &1  &10^-10^+&10^+10^-\\
 11       &     &3  &2  &3  &1  &11^+&11^- \\
 12       &2    &1  &1  &1  &1  &(12^-)^212^+&(12^+)^212^-\\
 13       &1    &1  &1  &2  &1  &13^-13^+&13^+13^-\\
 14       &1    &2  &2  &1  &1  &14^-14^+&14^+14^-\\
 15       &1    &2  &1  &2  &2  &15^-(15^+)^2&15^+(15-)^2\\
 16       &1    &2  &2  &2  &1  &16^-16^+&16^+16^-\\
 17       &1    &2  &2  &3  &1  &17^-17^+&17^+17^-\\
 18       &2    &3  &2  &2  &2  &(18^-)^2(18^+)^2&(18^+)^2(18^-)^2\\
 19       &1    &3  &2  &3  &2  &19^-(19^+)^2&19^+(19^-)^2\\
 20       &2    &3  &3  &3  &1  &(20^-)^220^+&(20^+)^2 20^-\\
 21       &2    &3  &2  &4  &2  &(21^-)^2(21^+)^2&(21^+)^2(21^-)^2\\
 22       &2    &4  &3  &3  &2  &(22^+)^2(22^-)^2&(22^-)^2(22^+)^2\\
 23       &1    &4  &3  &4  &2  &23^-(23^+)^2&23^+(23^-)^2\\
 24       &3    &2  &2  &2  &2  &(24^-)^3(24^+)^3&(24^+)^2(24^-)^2\\ \hline
 \end{array}
$
\caption{Multiplicity  of orbital partitions from  [KM66] for given values $0\leq \Lambda \leq 24$.\label{table1}}
\end{table}

The orbital partitions $[31],[4]$ do not apply  for electrons with spin $s=\frac{1}{2}$.
The entries for higher values of $\Lambda$ than given in Table \ref{table1} are found  by the recursive methods from \cite{KM66} eq. (6.8).

\section{States of four electrons.}

A full orbital 4-particle state is characterized by  the quantum numbers 
\begin{equation}
|\alpha^4 \rangle \equiv |\alpha^4 N, [h_1h_2], \Lambda, L,  f r\rangle, 
\end{equation}
where $r$ ranges over the orbital Young tableaus of the diagram  $f$.
The total spin $S$ is determined by antisymmetry and by the orbital partition $f$, see section 2.

\subsection{Four electrons, S=2.}
The orbital state has partition $f=[1111]$ of dimension $|[1111]|=1$. The operator eq.~\ref{es13}
applied to the initial state $|\alpha^4\rangle$ becomes 
\begin{equation}
 \label{es25}
|\alpha^4, [1111] \rangle = \sum_p T(p) |\alpha^4\rangle (-1)^p,
\end{equation}
where $(-1)^p$ is the sign of the permutation $p$, $+1$ for even and $-1$ for odd permutations.

\subsection{Four electrons, S=1.}
The orbital partition is $f=[211]$ of dimension $|[211]|=3$. 
The three Young diagrams (written for simplicity in rectangular frames) read
\begin{equation}
\label{es26}
\begin{array}{lll}
r_1=\left[
\begin{array}{ll}
1& 2\\
3&\\
4&\\
\end{array}
\right],
& r_2=\left[
\begin{array}{ll}
1& 3\\
2&\\
4&\\
\end{array}
\right],
& r_3=\left[
\begin{array}{ll}
1& 4\\
2&\\
3&\\
\end{array}
\right]\\
\end{array}.
\end{equation}
Assume that the initial state $|\alpha^4[211] r_1\rangle$ has been generated by application of 
a Young operator eq.\ref{es13}.
We find the  two partner states of $|\alpha^4[211]r_1\rangle$ by the operator relations eq \ref{es14} and the  axial distances,
\begin{eqnarray}
 \label{es27}
&|\alpha^4[211] r_2\rangle = |\alpha^4[211]r_1\rangle  \sqrt{\frac{1}{3}}
          +T(2,3)|\alpha^4[211]r_1\rangle \sqrt{\frac{4}{3}},\:\: \tau_{23}(r_1)=-2,\\
\nonumber           
&|\alpha^4[211] r_3\rangle = |\alpha^4[211]r_2\rangle \sqrt{\frac{1}{8}}
          +T(3,4)|\alpha^4[211]r_2\rangle \sqrt{\frac{9}{8}},\:\: \tau_{34}(r_2)= -3.
\end{eqnarray}
\subsection{Four electrons, S=0.}
The orbital partition is $f=[22]$ of dimension $|[22]|=2$. 
The two Young diagrams read
\begin{equation}
\label{es28}
r_1=\left[
\begin{array}{ll}
1&2\\
3&4\\
\end{array}
\right],
 r_2=\left[
\begin{array}{ll}
1&3\\
2&4\\
\end{array}
\right].
\end{equation}
Again we generate the initial orbital state $|\alpha^4[22]r_1\rangle$ by application of a Young operator eq.~\ref{es13}.
We find the second state by the   operator relation and the  axial distance
\begin{equation}
 \label{es29}
|\alpha^4[22] r_2\rangle = |\alpha^4[22]r_1\rangle \sqrt{\frac{1}{3}}
          +T(2,3)|\alpha^4[22]r_1\rangle \sqrt{\frac{4}{3}},\:\: \tau_{23}(r_1)=-2.
\end{equation}

\section{Relative tetrahedral coordinates for $n=4$ electrons.}
Since we are dealing with particle-particle interactions, the {\em cm} vector is not affected, the total momentum is conserved, and the system is amenable to a description in relative coordinates. 
A standard choice are the Jacobi relative coordinates
\begin{eqnarray}
\label{es29b}
&&\dot{\eta}=Jz,
\\ \nonumber
&&\dot{\eta}_1= \sqrt{\frac{1}{2}}(z_1-z_2),
\\ \nonumber 
&&\dot{\eta}_2=  \sqrt{\frac{1}{6}}(z_1+z_2)-2z_3).
\\ \nonumber
&&\dot{\eta}_3= \sqrt{\frac{1}{12}}(z_1+z_2+z_3)-3z_4).
\end{eqnarray}

In \cite{KM66} there was used a particular tetrahedral set of three doubledot relative coordinates. This set has the virtue that under the application of any permutation these relative coordinates are at most permuted, eq.~\ref{es31}.
The transformation properties of the doubledot coordinates are summarized in \cite{KM66} Table 3.
The relative tetrahedral coordinates $\ddot{\eta}$ from \cite{KM66}, but now enumerated in the order of \cite{GU13} for $n=4$, are 

\begin{eqnarray}
\label{es30}
&&\eta= Q z,
\\ \nonumber
&&\eta_1= \frac{1}{2}(z_1+z_2-z_3-z_4)=\ddot{\eta}_3,
\\ \nonumber 
&&\eta_2= \frac{1}{2}(z_1-z_2+z_3-z_4)=\ddot{\eta}_1,
\\ \nonumber 
&&\eta_3= \frac{1}{2}(z_1-z_2-z_3+z_4)=\ddot{\eta}_2.
\\ \nonumber 
\end{eqnarray}
It is found from \cite{KM66}, Table~3, that the tetrahedral coordinates eq.~\ref{es30} have extremely simple matrix transformations
under the generating transpositions $(1,2),(2,3),(3,4)$: 
\begin{eqnarray}
\label{es31}
&&T^{\eta}(1,2)\left[\begin{array}{r}
      \eta_1\\
      \eta_2\\
      \eta_3\\
      \end{array} \right]
= \left[\begin{array}{rrr}
  1&0&0\\
  0&0&-1\\
  0&-1&0\\
  \end{array} \right]
\left[\begin{array}{l}
      \eta_1\\
      \eta_2\\
      \eta_3\\
      \end{array}\right],
\\ \nonumber 
&&T^{\eta}(2,3)\left[\begin{array}{l}
      \eta_1\\
      \eta_2\\
      \eta_3\\
      \end{array} \right]
= \left[\begin{array}{lll}
  0&1&0\\
  1&0&0\\
  0&0&1\\
  \end{array} \right]
\left[\begin{array}{l}
      \eta_1\\
      \eta_2\\
      \eta_3\\
      \end{array}\right],
\\ \nonumber 
&&T^{\eta}(3,4)\left[\begin{array}{l}
      \eta_1\\
      \eta_2\\
      \eta_3\\
      \end{array} \right]
= \left[\begin{array}{lll}
  1&0&0\\
  0&0&1\\
  0&1&0\\
  \end{array} \right]
\left[\begin{array}{l}
      \eta_1\\
      \eta_2\\
      \eta_3\\
      \end{array}\right].
\end{eqnarray}

The relative coordinates in  Gusev \cite{GU13}  extend the tetrahedral coordinates eq.~\ref{es30}   
to $n>4$ particles. See Appendix B for their transforms under permutations.

We give now a first  example of the polynomial state construction in the tetrahedral coordinates eq.~\ref{es30} for the lowest excitation $N=3, [h_1h_2]=[30],  \Lambda=3, \: f=[211]$ with spin $S=1$ and the orbital Young tableaus $r_1,r_2,r_3$ of eq.~\ref{es16}: 
\begin{eqnarray}
\label{es32}
|\alpha^4[211]r_1\rangle&=&2\sqrt{\frac{2}{3}}(\eta_2-\eta_3)(\eta_2\eta_3+\eta_1\eta_1),
\\ \nonumber
|\alpha^4[211]r_2\rangle&=&\frac{2}{3}\sqrt{2}(\eta_2+\eta_3)(-\eta_1\eta_1+\eta_2\eta_3-2\eta_1(\eta_2-\eta_3)),
\\ \nonumber
|\alpha^4[211]r_3\rangle&=&\frac{4}{3}(\eta_2+\eta_3)(\eta_1-\eta_2)(\eta_3+\eta_1).
\end{eqnarray}

\section{From matrix elements of the Coulomb interaction to the  energy level structure.}

With the polynomial states for a specific set of quantum numbers at hand, we proceed to evaluate the various
parts of the four-electron Hamiltonian in the presence of a magnetic field $B$ perpendicular to the $(x,y)$ plane,
\begin{equation}
H=\sum_{i=1}^4 
\frac{p_{x,i}^2+p_{y,i}^2}{2m^*}
+\frac{1}{2}m^*\Omega^2 ({x_i}^2+{y_i}^2)
-\omega_{L} L_{z,i}
+\sum_{i<j} \frac{e^2}{4\pi\epsilon_0\epsilon \sqrt{{(x_i-x_j)}^2+{(y_i-y_j)}^2}},
\end{equation}
with the Larmor frequency $\omega_L=e B/(2m^*)$, the effective mass $m^*$ and the combined frequency $\Omega^2=\omega_0^2+\omega_L^2$ due to an external confinement potential $V(x,y)=\frac{1}{2}m^*\omega_0^2(x^2+y^2)$.
In all following expressions we work within the symmetric gauge, which allows one to express the uniform magnetic field in terms of two-dimensional oscillator excitations \cite{Kohn1961,Avron1978}.
The resulting four particle Hamiltonian is readily decomposed into  {\em cm} and relative parts in the plane perpendicular to the magnetic field direction. Without the Coulomb part it resembles the addition of 2D oscillator systems with frequencies $\Omega$.
The magnetic field induces an angular-momentum dependent energy shift and the non-interacting eigenenergies read
\begin{eqnarray}
E_{\rm cm}(N_{\rm cm},M_{\rm cm})&=&\hbar\Omega(2N_{\rm cm}+1+|M_{\rm cm}|)-\hbar\omega_L M_{\rm cm},\\
&&N_{\rm cm}=0,1,2,\ldots, M_{\rm cm}=0,\pm 1,\pm 2,\ldots\\
E_{\rm rel}(n_{\rm rel},m_{\rm rel})&=&\sum_{i=1}^3\hbar\Omega(2n_{i,\rm rel}+1+|m_{i,\rm rel}|)-\hbar\omega_L m_{i,\rm rel}\\
&&n_{\rm rel}=0,1,2,\ldots, m_{\rm rel}=0,\pm 1,\pm 2,\ldots
\end{eqnarray}
where we followed Fock's notation. In terms of the excitations $n_i^{\pm}$ these quantum-numbers read 
\begin{equation}
m_{i,\rm rel}=n_i^+ - n_i^-, \quad 2n_i=n_i^+ + n_i^- - |n_i^+ - n_i^-|
\end{equation}
The total energy is given by the addition of {\em cm} and relative parts
\begin{equation}
\label{es37}
E_{\rm tot}=
E_{\rm cm}(N_{\rm cm},M_{\rm cm})
+E_{\rm rel}(n_{\rm rel},m_{\rm rel})
+E_{\rm rel, Coulomb}.
\end{equation}
For the evaluation of the Coulomb interaction we express the states as polynomials in the Jacobi relative vectors.
There is, independent of the $cm$ coordinate,  a direct  linear matrix transformation from relative tetrahedral 
to Jacobi vectors obtained from 
eqs. \ref{es29b}, \ref{es30}, which for short we denote as
\begin{equation}
\label{es32b}
\dot{\eta} = J Q^{-1} \eta.
\end{equation}
The  first Jacobi vector from eq.~\ref{es29b} is the relative vector between particles 1,2
and according to eq.~\ref{es13e} it suffices to compute the full Coulomb interaction.

In the relative-coordinate polynomials the dimensionless Coulomb interaction reads
\begin{equation}
V_{\rm Coulomb}=6\frac{\kappa}{\sqrt{2}\dot{\eta}},\quad
\kappa=\frac{e^2}{4\pi\epsilon_0\epsilon} \frac{m^*}{\hbar^2} \sqrt{\frac{\hbar}{m^*\Omega}},
\end{equation}
where the factor $4(4-1)/2=6$ arises from the number of pairs for the two-body operator in eq.~\ref{es13e}.

The Coulomb matrix element between two states expressed in the three Jacobi
vectors becomes
\begin{eqnarray}
\label{es34}
&\langle n^+_1n^-_1n^+_2n^-_2n^+_3n^-_3|V_{\rm Coulomb}| m^+_1m^-_1m^+_2m^-_2m^+_3m^-_3 \rangle\nonumber\\
&=
\delta(n^+_2, m^+_2)
\delta(n^+_3, m^+_3)
\delta(n^-_2, m^-_2)
\delta(n^-_3, m^-_3)
\delta(\sum_{i=1}^3 (n^+_i-n^-_i),\sum_{i=1}^3 (m^+_i-m^-_i))\nonumber\\
&\times\delta(n^+_1 - n^-_1, m^+_1 - m^-_1)
V_{\rm C}({\rm min}[n^+_1, n^-_1], {\rm min}[m^+_1, m^-_1], |n^-_1 - n^+_1|),
\end{eqnarray} 
with  
\begin{eqnarray}
\label{es35}
V_{\rm C}(n,n',m)&=\frac{\kappa \pi {(-1)}^{n+n'} \Gamma \left(\left| m\right| +\frac{1}{2}\right)}{\sqrt{2}\Gamma(1/2-n)\Gamma(1+n)\Gamma(1/2-n')\Gamma(1+n')}
   \sqrt{\frac{n! }{(\left| m\right|+n )!}} 
   \sqrt{\frac{n'!}{(\left| m\right|+n')!}} \nonumber\\
& _3F_2\left(-n,-n',\left| m\right|
   +\frac{1}{2};\frac{1}{2}-n,\frac{1}{2}-n';1\right)
\end{eqnarray}

We construct for a given set of relative-coordinate excitations all relative-coordinate states $q_i$ and express each state 
$q_i(\dot{\eta}_1,\dot{\eta}_2,\dot{\eta}_3)$ as sum of polynomials in Jacobi coordinates with exponents
$n_{i,j}^\pm$
\begin{equation}
q_i=\sum_{j=1} a_{i,j} 
\dot{\eta}_1^{n_{1,i,j}^+}\dot{\eta}_1^{n_{1,i,j}^-} 
\dot{\eta}_2^{n_{2,i,j}^+}\dot{\eta}_2^{n_{2,i,j}^-} 
\dot{\eta}_3^{n_{3,i,j}^+}\dot{\eta}_3^{n_{3,i,j}^-}.
\end{equation}
Then the interaction matrix is given by
\begin{equation}
M_{\rm C}(q_i,q_k)=\sum_{j,m}a_{i,j}a^*_{k,m} 
\langle 
n_{1,i,j}^+ n_{1,i,j}^- n_{2,i,j}^+ n_{2,i,j}^- n_{3,i,j}^+ n_{3,i,j}^- |V_{\rm C}| 
n_{1,k,m}^+ n_{1,k,m}^- n_{2,k,m}^+ n_{2,k,m}^- n_{3,k,m}^+ n_{3,k,m}^- \rangle.
\end{equation}

The Coulomb interaction does not connect states with different angular momenta and 
so the blocks of states $(S,f)$ fall into subblocks labelled by $(S,f,m_{\rm rel})$.
However, different excitations of the relative oscillator are coupled by the Coulomb interaction.
The addition of the relative oscillators, including the Coulomb interaction is given by
diagonalization of the matrix
 \begin{equation}
M_{\rm rel}(q_i,q_k)=M_{\rm C}(q_i,q_k)+\delta_{ik}
\hbar\Omega\sum_{i=1}^3(1+n_i^+ + n_i^-)
\end{equation}
If we augment the relative eigenvalues with the analytic results for the {\em cm} energies,
we are in a position to compare the results with the configuration-interaction methods using 
Slater determinantes of single-particle states.

\subsection{Spin $S=2$ and the Laughlin state.}

\begin{table}
\begin{center}
\begin{tabular}{cccc}
 {\bf 1.8466} & 2.1596 & 2.3833 & 2.5451 \\
 {\bf 2.0012} & {\bf 2.2136} & {\bf 2.3899} & 2.6327 \\
 2.0217 & 2.2529 & 2.4127 & 2.6597 \\
 {\bf 2.0284} & 2.2589 & 2.4453 & 2.7362 \\
 2.0704 & 2.2959 & 2.4534 & 2.7974 \\
 2.1042 & {\bf 2.3132} & 2.4722 & 2.9591 \\
 2.1223 & 2.3133 & 2.4997 & 3.1498 \\
 {\bf 2.1265} & 2.3215 & 2.5245 &  \\
 2.1531 & 2.3832 & 2.5441 & \\
\end{tabular}
\end{center}
\caption{List of 34 SACI Coulomb eigenvalues ($\kappa=1$) in units of $\hbar\Omega$ for $S=2$ obtained by diagonalization of
the Slater determinants of the 34 single-particle product states ${\prod_{i=1}^4|\phi(0,l_i)\rangle}$ 
yielding $L=\sum_{i=1}^4 l_i=18$ within the lowest Landau level.
The entries in boldface indicate the seven states with the {\em cm} oscillator in the ground state
obtained by the polynomial method described in the text for the lowest Landau level ($n_{\rm rel}=m_{\rm rel}=18$),
while the remaining 27 eigenvalues correspond to spurious states.
}
\end{table}

As first example we discuss four electrons in a magnetic field with $\omega_0=0$ and the maximum  spin state $S=2$.
Due to the maximum spin state with all spins coinciding, the orbital state must be completely antisymmetric, $f=1^4$.
We choose $n_{\rm rel}=18$, $m_{\rm rel}=18$ (corresponding to the lowest Landau level) and the basis set consists of seven states with pseudo angular momenta
$\Lambda=6,10,12,14,16,{(18)}^2$.
The corresponding SACI basis set restricted to the lowest Landau level contains already additional 27 spurious states
due to the admixing of {\em cm} excitations.

It is instructive to see which linear combination of the seven non-spurious states yields the Laughlin state, which is
also free of any {\em cm} excitations.
The Laughlin antisymmetric trial function at filling factor 1/3 is written in terms of the 
single-particle states \cite{Laughlin1983a} 
\begin{equation}
\label{es36}
\psi_{1/3}={[(z^+_1-z^+_2)(z^+_1-z^+_3)(z^+_1-z^+_4)(z^+_2-z^+_3)(z^+_2-z^+_4)(z^+_3-z^+_4)]}^3.
\end{equation}
The Coulomb interaction with $\kappa=1$ yields an expectation value of $1.8535\,\hbar\Omega$ for the $\psi_{1/3}$ state,
slightly above the lowest eigenvalue $1.8466\,\hbar\Omega$ obtained by diagonalization of the Coulomb matrix $M_C$  
within the seven state basis set.

\subsection{Intrusion of {\em cm} excitation for the spin $S=0$ case.}

\begin{figure}[t!]
\begin{center}
\includegraphics[width=0.8\textwidth]{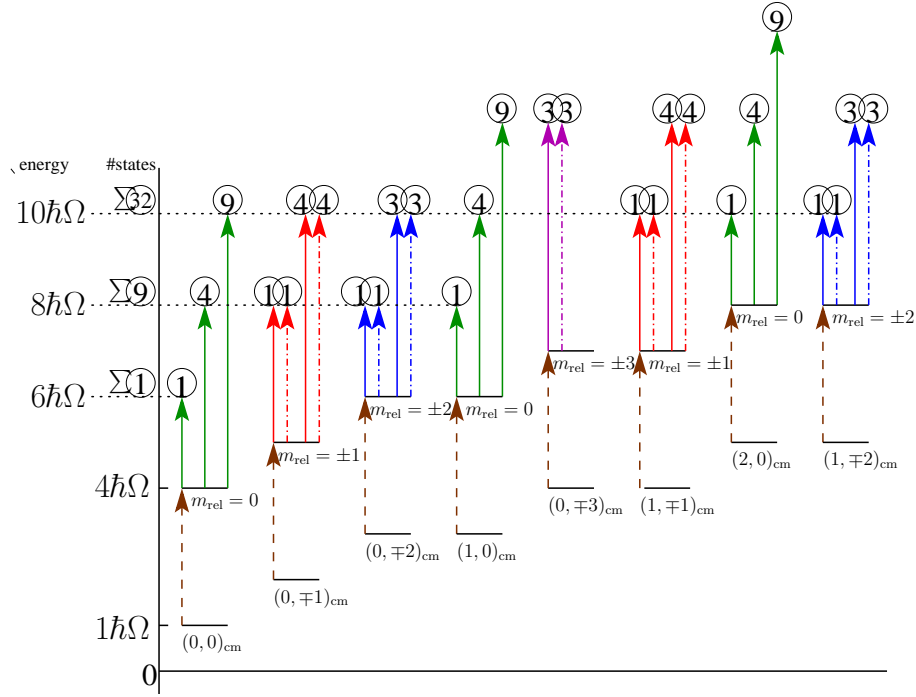}
\end{center}
\caption{Decomposition of the total angular state $L_{\rm tot}=0$ for spin $S=0$ for four electrons into relative-coordinate building
blocks. Excitations of {\em cm} angular momentum $M_{\rm cm}$ have to be combined with $m_{\rm rel}=-M_{\rm cm}$
values, which leads to a redundancy of interaction matrix elements. For instance, the 32 states at total energy 
$E=10\hbar\Omega$ can be broken down into three copies of relative-coordinate states (color coded). 
Each ladder starts with a {\em cm} state denoted by $(N_{\rm cm},M_{\rm cm})_{\rm cm}$, to which the zero-point 
energy $3$ of the relative-coordinate oscillators is added. The numbers in circles denote the number of relative-coordinate states with an excitation corresponding to the length of each arrow.
}\label{fig:levels0}
\end{figure}

As second example, we consider the case $f=[22]$ with total angular momentum $L_{\rm tot}=M_{\rm cm}+m_{\rm rel}=0$,
$\omega_L=0$. 
The relative-coordinate method allows one to decompose the problem into irreducible blocks which are separately diagonalized.
In Fig.~\ref{fig:levels0} the possible combinations of relative and {\em cm} angular momenta and excitations are depicted, which result in $L_{\rm tot}=0$.
If we consider all basis states with $E_{\rm total}^{\kappa=0}\le 10\hbar$ 
the Slater-determinantal SACI methods yields 42 basis states.
The irreducible group-theoretical approach yields decomposes these states into a block-diagonal basis sets of size 14 ($m_{\rm rel}=0$), 5 ($m_{\rm rel}=\pm 1$), and 4 ($m_{\rm rel}=\pm 2$), marked by green, red and blue arrows in Fig.~\ref{fig:levels0}. 
The additional 19 states (8 non-degenerate) of the SACI basis are all spurious states with $N_{\rm cm}>0$.  

\begin{figure}[t!]
\begin{center}
\includegraphics[width=0.7\textwidth]{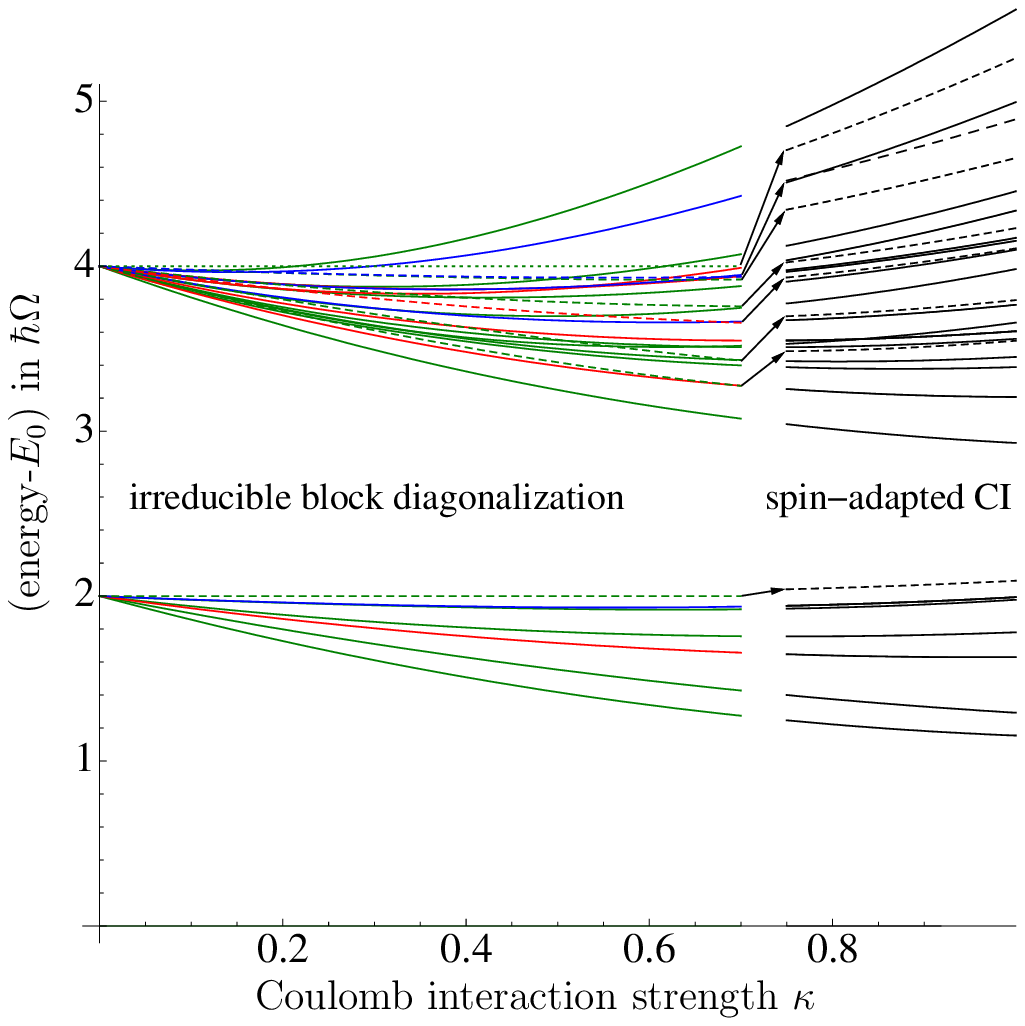}
\end{center}
\caption{Evolution of the lowest 42 eigenenergies of a four electron quantum dot ($\omega_L=0$) as function of the Coulomb interaction strength $\kappa$. The lowest eigenvalue $E_0$ has been subtracted.
The left curves  ($\kappa\le 0.7$) are obtained by the irreducible approach using relative-coordinates
and provide better converged values than the the spin-adapted configuration interaction method (SACI) (right curves $\kappa>0.75$).
The inaccuracies of the determinantal SACI method (upward arrows) affect all spurious states with $N_{\rm cm}>0$ (dashed and dotted lines).
}\label{fig:comp}
\end{figure}

Besides the smaller block sizes, another observation is that by fixing a cut-off energy in terms of 
$E_{\rm total}^{\kappa=0}$ in the configuration-interaction method, copies of basis elements with the same relative excitation and angular momentum are included but with varying number of excitations. 
For instance, while the $m_{\rm rel}=0$ state is included in the energy levels $E_{\rm total}^{\kappa=0} \le 10\hbar$ with three excitation levels $N=2,4,6$ on top of the $(N_{\rm cm}=0,M_{\rm cm}=0)_{\rm cm}$ state, it enters for $(1,0)_{\rm cm}$ with only 2 excited levels and at $(2,0)_{\rm cm}$ only with its ground state.

Fig.~\ref{fig:comp} demonstrates how the uneven inclusion of excited relative-coordinate states in the SACI basis results in a skewed eigenvalue spectrum of the SACI method and does not allow one to identify {\em cm} copies by scanning for eigenvalues spaced by the {\em cm} excitation $\hbar\Omega$.
In contrast, the separation of relative and {\em cm} states in the group-theoretical approach preserves
the exact spacing of the {\em cm} sets and and the diagonalization only affects the subspace of relative-coordinate states,
as required. Similar results apply to the spin $S=1,2$ states.

In particular SACI yields less-accurate eigenvalues at the higher eigenenergies (see the upward pointing arrows in Fig.~\ref{fig:comp}), while the group-theoretical approach
preserves the accuracy by combining the low-lying relative coordinate eigenvalues with the exact {\em cm} excitation energy.
Table~\ref{tab:convergence} gives the lowest eigenenergy for the $m_{\rm rel}=0$ state as function of relative coordinate excitation, which approaches the exact ground state eigenenergy with increasing basis size.

\begin{table}
\begin{tabular}{l|r|r|r|r|r}
maximal relative excitation & 2 &  4 &  6 &   8 &  10 \\\hline
basis size irred.\ rep.\    & 1 &  5 & 14 &  35 &  74 \\\hline
SACI basis size             & 1 & 10 & 42 & 145 & 405 \\\hline
lowest eigenenergy           & 11.483 & 10.685 & 10.592 & 10.555 & 10.539
\end{tabular}
\caption{Lowest eigenenergy for the $m_{\rm rel}=0$ case, Coulomb interaction $\kappa=1$, $\omega_L=0$ in units of $\hbar\Omega$. 
For the lowest eigenvalue, SACI and the irreducible block diagonalization yield identical results upon inclusion of
the same maximal relative excitation, however the SACI basis set is substantially larger due to the inclusion of spurious states.
\label{tab:convergence}}
\end{table}

\section{Summary of the group theoretical approach to construct the {\em cm} free basis.}

The construction of the basis set of relative coordinate polynomials for $n$ electrons entailed the following steps:\\

(1) Spin, partition: Given the number $n$ of electrons, we determine the possible spins $S$,  orbital partitions $f$ and angular momenta $L$. The 
state space falls into independent and orthogonal subspaces  labeled by $(S,f,L)$. Steps (1-7) work for n=4. 
For $n>4$ and steps (4-7) an extended analysis needs to be developed, some results are given in the appendices.

(2) Unitary quantum numbers: We provide 
the orbital quantum numbers of the system. For  a maximal relative 
oscillator excitation $N$ we list the unitary partitions $[h_1h_2], h_1+h_2=N$ for the  groups $U(2)\times U(n-1)$. 
A restriction to the lowest Landau level $LLL$ with $[h_1h_2] \rightarrow [N0]$ greatly simplifies the analysis.

(3) Angular momenta: The unitary partition $[h_1,h_2]$ from $U(2)>O(2)$ fixes the ladder of angular momenta $L$ from the maximum  value $L_{max}=h_1-h_2$ in steps of $2$ 
down to $0$ or $1$ and provides the final blocks $(S,f,L)$.

(4) Pseudo angular momentum and partition: The pseudo angular momentum $\Lambda$ is obtained from $[h_1,h_2]$ by the representation theory of the rotation subgroup in $O(n-1)< U(n-1)$.  
For $n=4$, any fixed pseudo angular momentum $\Lambda$ yields from \cite{KM66} p. 260 a precise list of  possible orbital partitions $f$. Multiple values of pseudo angular momentum yield linearly independent polynomials.

(5) State polynomials: Given  this list  of quantum numbers, we construct state polynomials  in relative coordinates. The  
initial polynomial is known from \cite{BM60}.
The  component of pseudo angular momentum and of total angular momentum are  lowered independently.

(6) Young tableau: For the known orbital partitions $f$, by
applying  permutation operators we project  states with definite Young tableau.

(7) Coulomb interaction and energies: For state polynomials in a block $(S,f,L)$,  
we isolate the first Jacobi vector in  bra and a ket states,  
evaluate the matrix elements of the Coulomb interaction, determine the hamiltonian matrix, diagonalize it and find 
the eigenstates of relative motion.

\section{Conclusion.}

The study of the four-electron quantum-dot  with Coulomb interactions in the relative-coordinate basis 
elucidates and overcomes some of the  problems inherent in the commonly used configuration interaction methods
such as SACI \cite{SACI,SACI2}:
If the $cm$ excitations, as in determinantal state space,  cannot be separated, a large cloud of spurious $cm$ excited  states, augmented by  spurious angular momenta, intrudes the variational state space
and obscures the variational energy spectrum reflecting the Coulomb interaction.
Within SACI, higher eigenenergies are distorted by the Coulomb interaction and a loss in accuracy is seen,
which can be avoided by the separation of the relative and $cm$ coordinates.
This allows one to construct  polynomial states with labels from representations of unitary and rotation groups
and to diagonalize the Coulomb-interaction part in smaller subblocks.
Our approach  opens a clear view on  the energy spectrum as shown in Figs. 1 and 2 and preserves
the exact ladder spacing $h\Omega$ of {\em cm} copies of the relative coordinates variational subspace. 
By increasing the basis size with respect to the relative excitations, 
the eigenenergies and eigenvectors obtained by diagonalization converge toward the 
exact eigenstates of the system.

\subsection*{Acknowledgements}

TK has been supported by the Heisenberg program of the DFG (KR 2889/5).

\newpage
\section{Appendix A: Alternative state analysis by use  of the subgroup $D_{2d}< S(4)$.}
An independent view on orbital  basis states for $4$ electons under permutations arises   by use of the subgroup $D_{2d}<S(4)$  from \cite{KM66}.
This method can be extended to higher excitations. It allows in Tables~\ref{table8},\ref{table9} to identify states  of
permutational symmetry for any pseudo angular momentum almost without projection.
We use $(\lambda, \mu)$ to denote a standard state of pseudo angular momentum $\lambda$ with component $\mu$, $\kappa= 0,1$ for parity, and $\rho=\pm 1$ with the states defined in eq.~\ref{a1}.
As mentioned in \cite{KM66} p. 263, the group $D_{2d}$ is the largest subgroup of $S(4)$ that leaves invariant the double-dot $\ddot{e}_3$ axis up to a reflection. With the Frobenius multiplicity expression we assign from the characters of $D_{2d}$, Table~\ref{table6}, in Table~\ref{table7} the reduction of IR f of $S(4)$ to those of $D_{2d}$. The numbers $(\kappa \rho \epsilon) $ 
are taken from \cite{KM66} and used in Table~\ref{table6}. They determine the irreducible bases of representation of $D_{2d}$ but from Table~\ref{table7} also basis states of permutational symmetry.
The basis states $[31]_3$ and $[211]_3$ are unique, and so their partner states may be derived by the ladder procedure \ref{es14}. The basis functions for $f=[31],[211]$ here transform like the doubledot representation. By \cite{KM66} eqs.~(6.1, 6.9) we can always return from the doubledot basis to the Jacobi basis.  The basis states of $f=[22]$ can be separated from the bases of 
$f=[4]$ and $f=[1^4]$ by projection.

We fix for the moment the labels $\kappa, \epsilon$ and introduce  $\rho=\pm 1$ in eq.~\ref{a1}.
Motivated by   \cite{KM66}, but independent of it, we redefine   two new  linear combinations of spherical  harmonics \cite{ED57} (2.5.7) by
\begin{eqnarray}
 \label{a1}
 &&\rho=1: |1(\lambda,\mu)\rangle = g_{\mu}[|(\lambda,\mu)\rangle+|(\lambda,-\mu)\rangle (-)^{\lambda+\mu}],\: \mu \geq 0,\\
 \nonumber
 &&\rho=-1: |-1(\lambda,\mu)\rangle = i^{-1}g_{\mu}[|(\lambda,\mu)\rangle-|(\lambda,-\mu)\rangle (-)^{\lambda+\mu}],
\:  \mu>0. 
\end{eqnarray}
These states by consulting \cite{KM66} Table 3 may be shown to be  orthogonal  eigenstates of the permutation operator $T((1,4)(2,3))$ with eigenvalue $\rho= \pm 1$, which from \cite{KM66} Fig 1 reverses  the axis  $\ddot{e}_3$.
States characterized by $(\kappa, \epsilon,\rho)$ we show to belong to irreps of the subgroup $D_{2d}$.
For the proof we employ the class operators \cite{KM66}, which are group operators $T(g)$ summed over a class of equivalent group elements, and 
commute with any group operator.
The five class operators $Ki$ \cite{KM66} of the group $D_{2d}$, written as sums  of permutation operators
from \cite{KM66}, Table~4 are
\begin{eqnarray}
\label{a2}
 && K1=T(e),\\
 \nonumber
 && K2= T((3,4)(1,2)),\\
 \nonumber
 && K3= T(1,3,2,4)+T(4,2,3,1),\\
 \nonumber 
 &&K4= T((1,3)(2,4))+T((1,4)(2,3)),\\
 \nonumber
 &&K5= T(1,2)+T(3,4).
 \end{eqnarray}
By use of the representation matrices of permutations we express the class operators eq.~\ref{a2} in matrix form, Table \ref{table12}. The matrix forms depend on the chosen 
representation, and we make convenient choices: 
For the representations $D^{[31]}$ and  $D^{[211]}, D^{[211]}(i,j)= I D^{[31]}((i,j)$ we use the basis eq.~\ref{es30} of $\ddot{\eta}_j$,
for the representation $D^{[22]}$ the Young representation \cite{HA62} p. 226. 
By virtue of these choices we find:  All class operators eq.~\ref{a2} have diagonal 
representations, see Table \ref{table12}. This proves:\\
{\bf Prop}:  All   basis states of the irreps  of the subgroup $D_{2d}$ in the chosen representations are basis states of irreducible representations of the bigger group  $S(4)$.\\ 

This remarkable result allows  in Table~\ref{table7}, up to certain ambiguities,  to almost avoid the use and projection with  Young operators
for the bigger group $S(4)$. In Tables~\ref{table8},\ref{table9} we use it in relation with the 
full scheme of groups including $SU(3)>O(3,R)$ and subgroups to
assign orbital symmetry to the oscillator states.

\begin{table}
$
\begin{array}{|l|l|l|l|l|l|} \hline
irrep  & E   &S_4  &C_2  &2C_2'  &2\sigma_d\\ \hline
1:A_1  & 1   & 1   & 1   & 1     & 1       \\ \hline
2:A_2  & 1   & 1   & 1   &-1     &-1       \\ \hline
3:B_1  & 1   & -1  & 1   & 1     &-1       \\ \hline
4:B_2  & 1   &-1   & 1   & -1    & 1       \\ \hline
5:E    & 2   & 0   &-2   & 0     & 0       \\ \hline
\end{array}
$
\caption{Characters of the five IR $1,..,5$ of $D_{2d}$  for its five classes in  
the irreducible representations.\label{table6}}
\end{table}

\begin{table}
$
\begin{array}{|l|l|l|l|l|l|} \hline
{\rm class} 
&D^{[4]}(Ki)
& D^{[31]}(Ki)
&D^{[211]}(Ki)
&D^{[22]}(Ki)
&D^{[1^4]}(Ki)\\ \hline
K1    
& \begin{array}{|l|} \hline 
   1\\ \hline
  \end{array}  
& \begin{array}{|l|l|l|}  \hline 
          1&  &  \\  \hline             
           & 1& \\   \hline
           &  &1\\   \hline
        \end{array}
&  \begin{array}{|l|l|l|}  \hline       
          1&  &  \\   \hline           
           & 1& \\    \hline
           &  &1\\    \hline
    \end{array}
& \begin{array}{|l|l|} \hline 
          1&   \\ \hline 
          & 1\\ \hline 
   \end{array} 
& \begin{array}{|l|} \hline 
   1\\  \hline
  \end{array}  \\ \hline   
K2    
& \begin{array}{|l|} \hline 
   1\\ \hline
  \end{array}  
& \begin{array}{|l|l|l|}  \hline 
         -1&  &  \\  \hline             
           &-1& \\   \hline
           &  &1\\   \hline
        \end{array}
&  \begin{array}{|l|l|l|}  \hline       
         -1&  &  \\   \hline           
           &-1& \\    \hline
           &  &1\\    \hline
    \end{array}
& \begin{array}{|l|l|} \hline 
          1&   \\ \hline 
          & 1 \\ \hline 
   \end{array}
& \begin{array}{|l|} \hline 
   1\\ \hline
  \end{array}\\ \hline  
K3    
& \begin{array}{|l|} \hline 
   2\\ \hline
  \end{array}  
& \begin{array}{|l|l|l|}  \hline 
           &  &  \\  \hline             
           &  & \\   \hline
           &  &-2\\   \hline
        \end{array}
&  \begin{array}{|l|l|l|}  \hline       
           &  &  \\   \hline           
           &  & \\    \hline
           &  &2\\    \hline
    \end{array}
& \begin{array}{|l|l|} \hline 
          1&   \\ \hline 
          & -1\\ \hline 
   \end{array}  
& \begin{array}{|l|} \hline 
   -2\\ \hline
  \end{array}\\ \hline    
K4
& \begin{array}{|l|} \hline 
   2\\ \hline
  \end{array}  
& \begin{array}{|l|l|l|}  \hline 
           &  &  \\  \hline             
           &  & \\   \hline
           &  &-2\\   \hline
        \end{array}
&  \begin{array}{|l|l|l|}  \hline       
           &  &  \\   \hline           
           &  & \\    \hline
           &  &-2\\    \hline
    \end{array}
& \begin{array}{|l|l|} \hline 
          1&   \\ \hline 
          & 1\\ \hline 
   \end{array}  
& \begin{array}{|l|} \hline 
   2\\ \hline
  \end{array}\\ \hline  
K5
& \begin{array}{|l|} \hline 
   2\\ \hline
  \end{array}  
& \begin{array}{|l|l|l|}  \hline 
           &  &  \\  \hline             
           &  & \\   \hline
           &  &2\\   \hline
        \end{array}
&  \begin{array}{|l|l|l|}  \hline       
           &  &  \\   \hline           
           &  & \\    \hline
           &  &-2\\    \hline
    \end{array}
& \begin{array}{|l|l|} \hline 
          2&   \\ \hline 
          &-2\\ \hline 
   \end{array} 
& \begin{array}{|l|} \hline 
   -2\\ \hline
  \end{array}\\   \hline 
\end{array}  
$
\caption{Diagonal matrix  representations  $D^f(Ki)$ for
partitions
$f=[4],[31],[211],[22],[1^4]$,
non-zero entries only, of the class operators $K1,...,K5$ eq.~\ref{a2} of $D_{2d}<S(4)$.
\label{table12}
}
\end{table}

\begin{table}
$
\begin{array}{|l|l|l|l|} \hline
\kappa  &\epsilon      &\rho=1         &\rho=-1          \\ \hline
 0      &+1            &[22]_2 | [4]    &[211]_3          \\ \hline
 0      &-1            &[22]_1  |[1^4]  &[31]_3           \\ \hline
 0      &0             &[31]_1 | [211]_1&[31]_2 | [211]_2   \\ \hline
1       &+1            &[22]_1 | [1^4]  &[31]_3           \\ \hline
1       &-1            &[22]_2 | [4]    &[211]_3          \\ \hline
1       &0             &[211]_1| [31]_1&[211]_2 | [31]_2    \\ \hline
\end{array}
$
\caption{
Basis states for all orbital partitions $f$ of $S(4)$ in correspondence to irreducible representations 
of the group $D_{2d}$ indexed 
by the labels of parity $\kappa= 0,1$, eigenvalue  $\rho=1,-1$ eq.~\ref{a1}, and 
$\epsilon:=(i^{|\mu|}+i^{-|\mu}|)/2=1,0,-1 $, from caption Table~\ref{table9}, compare \cite{KM66} p.267, Table 5.\label{table7}}
\end{table}

Of two states separated  by $|$, one and only one can belong to  the listed tableau. 
The states $[4],[1^4]$ are identified as eigenstates under  the transposition $T(2,3)$ with eigenvalue  $\pm 1$ respectively.
If a state is not reproduced under $T(2,3)$, it necessarily  belongs to $f=[22]$ and spin $S=0$.
We conclude  that the states eq.~\ref{a1} yield all the bases of
the orbital Young tableaus.

\begin{table}
$
\begin{array}{|l|l|l|l|l|l|l|l|l|l|}    \hline
N^{\pi}&\Lambda^{\pi}&|\mu|&  \kappa &\epsilon& \rho=1     &\rho=-1     & (\lambda',\mu')& \Lambda^{\pi}& f\\ \hline \hline
0^+&0^+           &0    &0        &1       &[4]+   & -          &(0,0)           & 0^+          &[4]\\ \hline
1^-&1^-           &1    &1        &0       &[31]1+  &[31]2+      &(1,0)           &1^-           &[31]\\ \hline
   &              &0    &1        &1       &-       & [31]3+     &                &              & \\   \hline  
2^+&2^+           &2    &0        &-1      &[22]1+  &[31]3+      &(2,0)           &2^+           &[31][22]\\ \hline
   &              &1    &0        &0       &[31]1+  &[31]2+      &                &              &       \\ \hline
   &              &0    &0        &1       &[22]2+  &-           &                &              &        \\  \hline   
   &1^+           &1    &0        &0       &[211]1+ &[211]2+     &(0,1)           &1^+           & [211]  \\ \hline
   &              &0    &0        &1       &-           &[211]3+ &                &              &  \\ \hline
   &0^+           &0    &0        &1       &[4]+    & -          &(2,0)           &0^+           &[4]  \\ \hline 
3^-&3^-           &3    &1        &0       &[211]1|[31]1*&[211]2|[31]2*&(3,0)         &3^-           &[211][31][4]\\ \hline 
   &              &2    &1        &-1      &[4]+    &[211]3+     &                &              & \\ \hline
   &              &1    &1        &0       &[211]1|[31]1*&[211]2|[31]2*&              &              & \\ \hline
   &              &0    &1        &1       &-           &[31]3+  &                &              &  \\ \hline
   &2^-           &2    &1        &-1      &[22]2+  &[211]3+     &(1,1)           &2^-           &[211][22] \\ \hline
   &              &1    &1        &0       &[211]1+ &[211]2+     &                &              &      \\ \hline
   &              &0    &1        &1       &[22]1+  &-           &                &              &  \\ \hline
   &1^-           &1    &1        &0       &[31]1+  &[31]2+      &(3,0),(1,1)     &(1^-)^2       &[31]^2  \\ \hline
   &              &0    &1        &1       &-       &[31]3+      &                &              &   \\ \hline
4^+&4^+           &4    &0        &1       &[22]2|[4]*&[211]3*     &(4,0)           &4^+           &[4][31][22][211]\\ \hline
   &              &3    &0        &0       &[31]1|[211]1*&[31]|2[211]2*&                &              &   \\ \hline
   &              &2    &0        &-1      &[22]1+  &[31]3+&                &              &   \\ \hline
   &              &1    &0        &0       &[31]1|[211]1*&[31]2|[211]2*&                &              &   \\ \hline
   &              &0    &0        &1       &[22]2|[4]*   &[211]3*    &                &              &   \\ \hline
   &3^+           &3    &0        &0       &[31]1|[211]1*&[31]2|[211]2*&(2,1)           &3^+           &[31][211][1^4]\\ \hline
   &              &2    &0        &-1      &[1^4]       &[31]3+  &                &              &     \\ \hline
   &              &1    &0        &0       &[31]1|[211]1*&[31]2|[211]2*&                &              & \\ \hline
   &              &0    &0        &1       & -          &[211]3+ &                &              & \\ \hline
   &2^+           &2    &0        &-1      &[22]1+  &[31]3+      &(4,0),(2,1),(0,2)&(2^+)^3          &[31]^3[22]^3 \\ \hline
   &              &1    &0        &0       &[31]1+  &[31]2+      &                &              & \\ \hline
   &              &0    &0        &1       &[22]2+  &-           &                &              & \\ \hline
   &1^+           &1    &0        &0       &[211]1+&[211]2+      & (2,1)          &1^+           &[211]\\  \hline
   &              &0    &0        &1       &-       &[211]3+     &                &              & \\ \hline
   &0^+           &0    &0        &1       &[4]+    &-           &(4,0),(0,2)     &(0^+)^2           &[4]^2\\ \hline 
\end{array}
$
\caption{continued on next table\label{table8}}
\end{table}

\begin{table}
$
\begin{array}{|l|l|l|l|l|l|l|l|l|l|}    \hline
N^{\pi}& \Lambda^{\pi}&|\mu|      &  \kappa &\epsilon& \rho=1     &\rho=-1     & (\lambda',\mu')& \Lambda^{\pi}& f\\ \hline \hline
5^-&5^-               &5          &1        &0       &[211]1|[31]1*&[211]2|[31]2* &(5,0)           &5^-           &[211][22][31]^2\\ \hline      
   &                  &4          &1        &1       &[22]1+       &[31]3+     &                 &              &  \\ \hline
   &                  &3          &1        &0       &[211]1|[31]1*&[211]2|[31]2*&               &              & \\ \hline
   &                  &2          &1        &-1      &[22]2+       &[211]3+    &                 &              & \\ \hline
   &                  &1          &1        &0       &[211]1|[31]1*&[211]2|[31]2*&               &              &  \\ \hline
   &                  &0          &1        &1       &[22]1+       &[31]3+     &                 &              &  \\ \hline
   &4^-               &4          &1        &1       &[1^4]|[22]1* &[31]3+     &(3,1)            &4^-           &[1^4][211][22][31]\\ \hline
   &                  &3          &1        &0       &[211]1|[31]1*&[211]2|[31]2*&               &              &  \\ \hline
   &                  &2          &1        &-1      &[22]2+       &[211]3+     &                &              &  \\ \hline
   &                  &1          &1        &0       &[211]1|[31]1*&[211]2|[31]2*&               &              &  \\ \hline
   &                  &0          &1        &1       &[22]1|[1^4]* &[31]3+      &                &              &  \\ \hline
   &3^-               &3          &1        &0       &[211]1|[31]1*&[211]2|[31]2*&(5,0)(3,1)(1,2)&(3^-)^3           &[211]^3[31]^3[4]^3 \\ \hline  
   &                  &2          &1        &-1      &[4]+         &[211]3+    &                 &              &  \\ \hline
   &                  &1          &1        &0       &[211]1|[31]1*&[211]2|[31]2*&               &              &   \\ \hline
   &                  &0          &1        &1       &-            &[31]3+     &                 &              &   \\ \hline
   &2^-               &2          &1        &-1      &[22]2+       &[211]3+    &(3,1)(1,2)       &(2^-)^2           &[211]^2[22]^2  \\ \hline
   &                  &1          &1        &0       &[211]1+      &[211]2+    &                 &              &  \\ \hline
   &                  &0          &1        &1       &[22]1+       & -         &                 &              &    \\ \hline
   &1^-               &1          &1        &0       &[31]1+       &  [31]2+   &(5,0)(3,1)(1,2)  &(1^-)^3       &[31]^3 \\ \hline  
   &                  &0          &1        &1       &-           &[31]3+      &                 &              &  \\ \hline
\end{array}   
$
\caption{
Permutational states   combining  $D_{2d}$   with $SU(3,C)$ labels.\newline
$N\leq 5$ total excitation/degree, Orbital 2D angular momentum range: $L= \lambda', \lambda'-2,\ldots, 1\: {\rm or}\:0$.
States marked $+$ are unambigueous.
  $\Lambda$ pseudo angular momentum with component $\mu$, $\pi=\pm 1$  standard parity. \newline 
$D_{2d}$, columns 1-7 from \cite{KM66}:$|\mu|$ absolute value of $\Lambda's$ component, $\kappa= 0,1$ \cite{KM66} parity, 
$\rho=\pm 1$ eigenvalue eq.~\ref{a2}  of pseudo-spherical state eq.~\ref{a1}. Vertical bars separate ambigueous choices marked by *, select one by ladder procedure.
Note that labels depend only on $\Lambda,\kappa$. \newline 
$SU(3,C)$, columns 8-10 from \cite{MS96},\cite{KM66}: $(\lambda',\mu')$ standard irrep labels, $\Lambda^{\pi}$ pseudo angular momentum \cite{MS96} p.177 and parity $\pi$, $f$ orbital  partition of $S(4)$, multiplicity from \cite{KM66} p.260.}

\label{table9}

\end{table}

\newpage
\section{Appendix B: Symmetrized relative coordinates for $n>4$ electrons and their permutations.}

The efficiency  of the tetrahedral coordinates raises the question if similar relative coordinates exist for $n>4$.
As a generalization of the tetrahedral coordinates from \cite{KM66}, new symmetrized coordinates for $n$ particles were proposed by Gusev et al \cite{GU13}.
The matrix that gives the $n$ new coordinates $(\eta_0,\eta_1,..,\eta_{n-1})$ in terms of the old ones $(x_1,x_2,..,x_n)$ reads 
\begin{eqnarray}
 \label{b1}
&& C= \frac{1}{\sqrt{n}}
 \left[
 \begin{array}{lllllll}
 1&1&1&1&...&1&1\\
 1&b&a&a&...&a&a\\
 1&a&b&a&...&a&a\\
 .&.&.&.&...&.&.\\
 1&a&a&a&...&b&a\\
 1&a&a&a&...&a&b\\
 \end{array}
 \right],
 \\ \nonumber
 &&C=C^T,\: C^{-1}=C,\: C^2=I,
 \\ \nonumber
 && a=[1-\sqrt{n}]^{-1},\: b=a+ \sqrt{n}.
 \end{eqnarray}
The {\em cm} coordinate is included as $\eta_0$.
We shall explore the properties of these coordinates under the action of permutations.
All permutations are generated by products of the $n-1$  transpositions 
$(1,2),(2,3),..,(n-1,n)$. We need the action of these transpositions on the new relative coordinates.
We shall see in eq.~\ref{b4} that under permutations $h\in S(n-1)$ acting on particles $(2,3,..,n)$ 
the symmetrized coordinates transform like single particle coordinates. So we focus on the remaining action 
$D^{\eta}(1,2)$ of the first transposition $(1,2)$ with respect to  the new coordinates. 

\subsection{The example of  $n=5$ electrons.}
For  $n=5$ we obtain for the linear action of the transposition $(1,2)$ on the new coordinates
\begin{eqnarray}
 \label{b2}
&& D^{\eta}(1,2)=\frac{1}{5}
\left[
 \begin{array}{lllll}
 1&1&1&1&1\\
 1&b&a&a&a\\
 1&a&b&a&a\\
 1&a&a&b&a\\
 1&a&a&a&b\\
 \end{array}
 \right]
 \left[
 \begin{array}{lllll}
 0&1&0&0&0\\
 1&0&0&0&0\\
 0&0&1&0&0\\
 0&0&0&1&0\\
 0&0&0&0&1\\
 \end{array}
 \right] 
 \left[
 \begin{array}{lllll}
 1&1&1&1&1\\
 1&b&a&a&a\\
 1&a&b&a&a\\
 1&a&a&b&a\\
 1&a&a&a&b\\
 \end{array}
 \right] 
 \\ \nonumber
&&= \frac{1}{5}
\left[
 \begin{array}{l|l|l|l|l} \hline
 5     &1+3a+b      &1+3a+b     &1+3a+b     &1+3a+b\\ \hline
 1+3a+b&2b+3a^2     &a+b+2a^2+ab&a+b+2a^2+ab&a+b+2a^2+b^2\\ \hline
 1+3a+b&a+b+2a^2+ab &2a+2a^2+b^2&2a+a^2+2ab &2a+a^2+2ab\\ \hline
 1+3a+b&a+b+2a^2+ab &2a+a^2+2ab &2a+2a^2+b^2&2a+a^2+2ab\\ \hline
 1+3a+b&a+b+2a^2+b^2&2a+a^2+2ab &2a+a^2+2ab &2a+2a^2+b^2\\ \hline
 \end{array}
 \right] 
\end{eqnarray}
This matrix is symmetric and orthogonal by construction, its square is the unit matrix.

We compute from eqs.\ref{es21},\ref{es26},\ref{es30}   the numbers 
\begin{eqnarray}
\label{b3}
&& a=-\frac{1}{4}[1+\sqrt{5}],\: b= -\frac{1}{4}[1-3\sqrt{5}],
\\ \nonumber
&&(002300)=2b+3a^2=\frac{1}{8}[5-\sqrt{5}],
\\ \nonumber
&&(011201)=a+b+2a^2+ab=\frac{5}{8}[-1+\sqrt{5}],
\\ \nonumber
&&(020210)=2a+2a^2+b^2=\frac{5}{8}[5-\sqrt{5}],
\\ \nonumber 
&&(020102)=2a+a^2+2ab=-\frac{5}{8}[3+\sqrt{5}].
\end{eqnarray}
For $n=5$, all  the numbers and matrices can be decomposed into
a rational part and a second part rational but proportional to $\sqrt{5}$. All these numbers form a  module that  closes under  addition, multiplication and division, compare the example eq.~\ref{es23}. 
We give this decomposition in the following examples. A similar module decomposition applies for any  $n$ whose square root
is not an integer.
\begin{eqnarray}
\label{b4}
&& C= A+\sqrt{5}B,
\\ \nonumber 
&&A=\frac{1}{4}
\left[
\begin{array}{rrrrr}
0     &0 &0 &0 &0\\
0     &3 &-1&-1&-1\\
0     &-1&-1&-1&-1\\
0     &-1&-1&-1&-1\\
0     &-1&-1&-1 &3\\
\end{array}
\right],
B=\frac{1}{4\cdot 5}
\left[
\begin{array}{rrrrr}
4     &4 &4 &4 &4\\
4     &-1&-1&-1&-1\\
4     &-1&-1&-1&-1\\
4     &-1&-1&-1&-1\\
4     &-1 &-1&-1 &-1\\
\end{array}
\right],
\\ \nonumber
&&D^{\eta}(1,2)= A^{\eta}+ \sqrt{5}B^{\eta}, 
\\ \nonumber
&&A^{\eta}= \frac{1}{5\cdot8}
\left[
\begin{array}{rrrrr}
5\cdot 8&0 & 0& 0 & 0\\
0       &5     &-5&-5 &-5\\
0       &-5&25&-15&-15\\
0       &-5&-15&25&-15\\
0       &-5&-15&-15&25\\
\end{array}
\right],
\: B^{\eta}=  \frac{1}{5\cdot 8}
\left[
\begin{array}{rrrrr}
0      &0   &0   &0   &0\\
0      &-1   & 5 &5  &5\\
0      &5   &-5  &-5  &-5\\
0      &5   &-5  &-5  &-5\\
0      &5   &-5  &-5  &-5\\
\end{array}
\right].
\end{eqnarray}

We find with eq.\ref{b3} from  the values of $a,b$ for $n=5$,
\begin{equation}
\label{b5}
1+3a+b=0
\end{equation}
and so the product matrix eq.\ref{b2}   reduces to its diagonal $4 \times 4$ submatrix that involves only 
the relative coordinates $\eta_1,..,\eta_{n-1}$. This is always necessary, since the transposition $(1,2)$
cannot affect the {\em cm} coordinate $\eta_0$.
For the other transpositions $(i,i+1), i>1$ we find the result
\begin{equation}
\label{b6}
2\leq i\leq (n-1)=4: T(i,i+1)\eta_{i-1}=\eta_{i},\: T(i,i+1)\eta_{i}=\eta_{i-1} 
\end{equation}
So all transpositions $(i,i+1)$ in single particle coordinates $2,...,n$  act on the relative coordinates
$(\eta_1,\eta_2,..,\eta_{n-1})$ like  transpositions of relative coordinates which we denote as $T^{\eta}(i-1,i)$. 
We summarize eq.~\ref{b4} in symbolic form as 
\begin{equation}
\label{b7}
T(i,i+1) \sim T^{\eta}(i-1,i),\: i=2,..,n-1.
\end{equation}
The transposition $T(1,2)$, missing in eq.~\ref{b7}, and acting on the symmetrized coordinates, is given by the matrix $D^{\eta}(1,2)$ eq.~\ref{b10}. 

\subsection{Permutations of symmetrized relative coordinates for $n$ electrons.}
The results of the last subsection, in particular eq.~\ref{b7} generalize to any $n$. Since the transpositions eq.~\ref{b4} generate the subgroup $S(n-1)$, all permutations of particles $(2,..,n)$ are mapped by eq.~\ref{b7} into permutations of the relative coordinates $(\eta_1,...,\eta_{n-1})$. The correspondence eq.~\ref{b7} must be kept in mind when working with the new relative coordinates.

We could easily antisymmetrize 
a polynomial state wrt to the (n-1) particles $2,3,..,n$,  but must augment  the antisymmetrizer  $A^{n-1}$ to include particle $1$, see the next section.
In terms of the symmetric group, we have the subgroup $S(n-1)$ acting on particles $2,3,..,n$. All elements $p \in S(n)$ can  be written as $p=c_i h$ with  $h \in S(n-1)$, multiplied by the $n$ coset generators  which may be chosen 
as transpositions $c_0=e, c_i= (1,i), i=2,..,n$. These coset generators  can be rewritten  as 
\begin{equation}
\label{b8}
(1,i) = (i,2)(1,2)(2,i),\: i=3,..,n,  (i,2) \in S(n-1) 
\end{equation}
and so their  matrices $D^{\eta}(1,i)$  acting on relative coordinates reduce to the element $D^{\eta}(1,2)$ whose matrix we compute in eq.~\ref{b7}.
The representations $D^{\eta}(i,2), i=2,...,n$ appearing in eq.~\ref{b5} simply mean the 
interchange of two rows or two columns. So our main task is to give the generalization of eq.~\ref{b2} for any $n$.
We denote the matrix elements of $D^{\eta}(1,2)$ for short by six integers 
\begin{eqnarray}
 \label{b9}
&& (n_0n_1n_2n_3n_4n_5)\rightarrow (n_0 +n_1a+n_2b+n_3a^2+n_4b^2+n_5ab),
\\ \nonumber
&&a=[1-\sqrt{n}]^{-1}, b=a+ \sqrt{n},
\end{eqnarray}
and obtain for the matrix representation in $\eta$-coordinates and general $n$ the result
\begin{eqnarray}
 \label{b10}
&& D^{\eta}(1,2)= C D(1,2)C^{-1}
\\ \nonumber
&&
=\frac{1}{n}\left[
 \begin{array}{llllll}
 1&1&1&1&1&.\\
 1&b&a&a&a&.\\
 1&a&b&a&a&.\\
 1&a&a&b&a&.\\
 1&a&a&a&b&.\\
 .&.&.&.&.&.\\
 \end{array}
 \right]
 \left[
 \begin{array}{llllll}
 0&1&0&0&0&.\\
 1&0&0&0&0&.\\
 0&0&1&0&0&.\\
 0&0&0&1&0&.\\
 0&0&0&0&1&.\\
 .&.&.&.&.&.\\
 \end{array}
 \right] 
 \left[
 \begin{array}{llllll}
 1&1&1&1&1&.\\
 1&b&a&a&a&.\\
 1&a&b&a&a&.\\
 1&a&a&b&a&.\\
 1&a&a&a&b&.\\
 .&.&.&.&.&.\\
 \end{array}
 \right] 
 \\ \nonumber
&&= \frac{1}{n}
\left[
 \begin{array}{l|l|l|l|l|l} \hline
 (n00000) &(1n\!-\!21000) &(1n\!-\!21000) &(1n\!-\!21000) &(1n\!-\!21000)  &.\\ \hline
 (1n\!-\!21000)&(002n\!-\!200)&(011n\!-\!301)&(011n\!-\!301)&(011n\!-\!301)&.\\ \hline
 (1n\!-\!21000)&(011n\!-\!301)&(020n\!-\!310)&(020n\!-\!402)&(020n\!-\!402)&.\\ \hline
 (1n\!-\!21000)&(011n\!-\!301)&(020n\!-\!402)&(020n\!-\!310)&(020n\!-\!402)&.\\ \hline
 (1n\!-\!21000)&(011n\!-\!301)&(020n\!-\!402)&(020n\!-\!402)&(020n\!-\!310)&.\\ \hline
 ..             & ..            & ..            & ..            & ..             & .\\ \hline
 \end{array} \right]
\end{eqnarray}
The matrix elements are given in the notation eq.~\ref{b9}. From eq.~\ref{b10} the full  matrix $D^{\eta}(1,2)$  is constructed for any $n$: The 
numbers $a,b$ are inserted from eq.~\ref{b1} as functions of $n$. The diagonal elements indexed by $((i,i),  i\geq 3)$ 
are repeated along the diagonal,
the elements in the lines on top,  and columns below the diagonal, are repeated in rows to the right and in columns downwards respectively  by the same functions eq.~\ref{b9} of $n$.
Again $D^{\eta}(1,2)$ is orthogonal and symmetric, its square is the unit matrix.
We find in general from eq.~\ref{b1} for the entries of the first row and column of $D^{\eta}(1,2)$
\begin{equation}
 \label{b11}
(1n\!-\!21000)\rightarrow (1+(n-2)a+b)=0,
\end{equation}
and so the $(n\times n)$  matrix $D^{\eta}(1,2)$  reduces to its $(n-1)\times (n-1)$ diagonal submatrix acting exclusively on the relative coordinates
$(\eta_1,\eta_2,..,\eta_{n-1})$.
In building orbital states from the symmetrized Gusev coordinates we must take care in the application of 
permutation operators.

\end{document}